\documentclass[sigconf]{acmart}
\AtBeginDocument{%
  \providecommand\BibTeX{{%
    \normalfont B\kern-0.5em{\scshape i\kern-0.25em b}\kern-0.8em\TeX}}}

\copyrightyear{2023}
\acmYear{2023}
\setcopyright{acmlicensed}
\acmConference[MM '23] {Proceedings of the 31st ACM International Conference on Multimedia}{October 29--November 3, 2023}{Ottawa, ON, Canada.}
\acmBooktitle{Proceedings of the 31st ACM International Conference on Multimedia (MM '23), October 29--November 3, 2023, Ottawa, ON, Canada}
\acmPrice{15.00}
\acmISBN{979-8-4007-0108-5/23/10}
\acmDOI{10.1145/3581783.3612092}

\usepackage{balance}
\usepackage{algorithm}
\usepackage{algorithmic}
\usepackage{subfigure}
\usepackage{multirow}
\usepackage{bm}

\newcommand{\cready}[1]{#1}

\begin{document}

\title{Isolation and Induction: Training Robust Deep Neural Networks against Model Stealing Attacks}

\author{Jun Guo}
\orcid{0000-0002-6626-4135}
\affiliation{
  \institution{Beihang University, China}
  \country{}
}
\email{junguo@buaa.edu.cn}

\author{Aishan Liu}
\orcid{0000-0002-4224-1318}
\affiliation{
  \institution{\emph{NLSDE}, Beihang University, China}
  \institution{Institute of Dataspace, Hefei, China}
  \country{}
}
\email{liuaishan@buaa.edu.cn}
\authornote{The corresponding author.}

\author{Xingyu Zheng}
\orcid{0009-0009-6283-7635}
\affiliation{
  \institution{Beihang University, China}
  \country{}
}
\email{xingyuzheng@buaa.edu.cn}

\author{Siyuan Liang}
\orcid{0000-0002-6154-0233}
\affiliation{
  \institution{Chinese Academy of Sciences, China}
  \country{}
}
\email{liangsiyuan@iie.ac.cn}

\author{Yisong Xiao}
\orcid{0000-0001-8227-0052}
\affiliation{
  \institution{Beihang University, China}
  \country{}
}
\email{xiaoyisong@buaa.edu.cn}

\author{Yichao Wu}
\orcid{}
\affiliation{
  \institution{Sensetime Group Limited, China}
  \country{}
}
\email{wuyichao@sensetime.com}

\author{Xianglong Liu}
\orcid{0000-0002-7618-3275}
\affiliation{
  \institution{\emph{NLSDE}, Beihang University, China}
  \institution{Zhongguancun Laboratory, China} 
  \institution{Institute of Dataspace, Hefei, China}
  \country{}
}
\email{xlliu@buaa.edu.cn}


\renewcommand{\shortauthors}{Jun Guo et al.}

\begin{abstract}
      Despite the broad application of Machine Learning models as a Service (MLaaS), they are vulnerable to model stealing attacks. These attacks can replicate the model functionality by using the black-box query process without any prior knowledge of the target victim model. Existing stealing defenses add deceptive perturbations to the victim's posterior probabilities to mislead the attackers. However, these defenses are now suffering problems of high inference computational overheads and unfavorable trade-offs between benign accuracy and stealing robustness, which challenges the feasibility of deployed models in practice. To address the problems, this paper proposes \textit{Isolation and Induction} (InI), a novel and effective training framework for model stealing defenses. Instead of deploying auxiliary defense modules that introduce redundant inference time, InI directly trains a defensive model by isolating the adversary's training gradient from the expected gradient, which can effectively reduce the inference computational cost. In contrast to adding perturbations over model predictions that harm the benign accuracy, we train models to produce uninformative outputs against stealing queries, which can induce the adversary to extract little useful knowledge from victim models with minimal impact on the benign performance. Extensive experiments on several visual classification datasets (\textit{e.g.}, MNIST and CIFAR10) demonstrate the superior robustness (up to 48\% reduction on stealing accuracy) and speed (up to 25.4× faster) of our InI over other state-of-the-art methods. Our codes can be found in \url{https://github.com/DIG-Beihang/InI-Model-Stealing-Defense}.
\end{abstract}




\maketitle

\section{Introduction}
Machine learning (ML) models, especially deep neural networks (DNNs), have been deployed across a wide range of areas, especially in computer vision.
Currently, Machine Learning as a Service (MLaaS) platforms have emerged to outsource well-trained deep learning models for developers since it often requires high computational resources to build a sophisticated ML model. 

\begin{figure}[t]
\begin{center}
   \includegraphics[width=\linewidth]{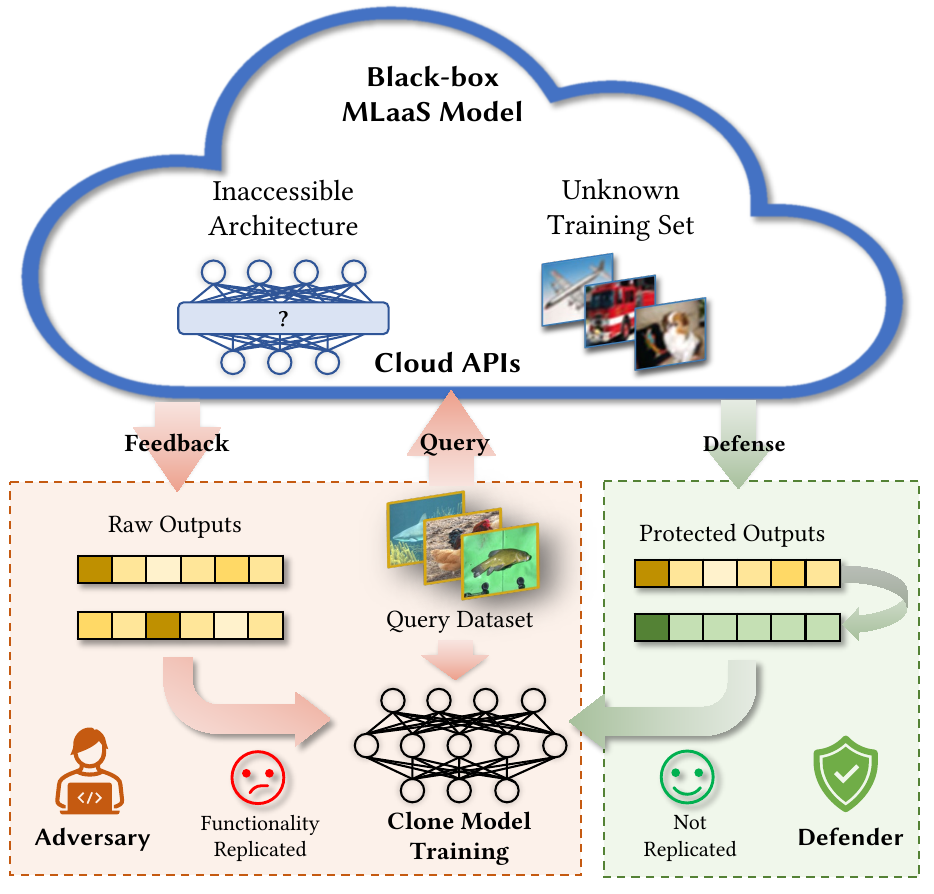}
\end{center}
   \caption{Illustration of attacks and defenses in model stealing. The adversary makes queries using malicious queries to extract knowledge from the victim MLaaS model, and the returned outputs are used to train a clone model. The defender introduces randomness into the model's outputs, in order to mislead the stealing algorithm.}
\label{fig:intro}
\end{figure}

However, severe security issues exist when using online platforms since the knowledge of these ML models is exposed to the risk of being stolen. Extensive studies have revealed that the functionality of an ML model can be extracted by adversaries, even when they have no knowledge about training examples or model structures~\cite{tramer2016stealing, orekondy2019knockoff, wallace2020imitation, kariyappa2021maze, yuan2022attack, liang2022imitated}. Adversaries can easily steal a deployed (victim) model by querying it using partial or surrogate input data, which is called \textit{model stealing} or \textit{model extraction} attacks. Model stealing attacks have raised increasing attention since they have posed strong security threats, where attackers can acquire a function-similar copy of the victim model to extract confidential data \cite{fredrikson2015model} or even perform adversarial attacks~\cite{szegedy2013intriguing, goodfellow2014explaining, liu2019perceptual, liu2020bias, liu2020spatiotemporal, wang2021dual, li2022hierarchical, liu2022harnessing, liu2023x, li2023towards, wei2018transferable, liang2020efficient, liang2022parallel, liang2022large, ma2022tale, ma2021poisoning} via the substitute for the victim model~\cite{papernot2017practical}. 

To mitigate the threat of model stealing attacks, several defensive methods~\cite{lee2019defending, orekondy2019prediction, mazeika2022steer, lee2022model} have been devoted to making the victim model hard to steal by introducing perturbations or randomness to the model output. However, the practical feasibility of these defenses is still hampered by certain limitations: (1) Existing defensive methods often incorporate auxiliary modules that validate the input or modify the output of the victim model, which introduces extra computational costs. In practice, the computational burdens are mainly concentrated on the inference phase, and higher computational overheads indicate extended user response time and increased financial expenditures. (2) Some defenses add perturbations to the model's output to enhance its resilience against model stealing attacks by providing erroneous predictions to attackers. However, this comes at the expense of reduced benign accuracy for legitimate users due to the unfavorable trade-off.

To tackle these concerns, we propose a novel and effective defensive training framework against model stealing attacks, which is called InI. As for the \emph{computational overheads}, distinct from prevailing methodologies that introduce auxiliary inference-time modules, our InI aims to directly train a robust model that is able to defend against stealing attacks without extra inference modules. Based on the fact that DNNs are heavily over-parameterized~\cite{zhang2017understanding} and can be trained to fit and generalize across diverse data distributions, we, therefore, posit that the victim model can achieve robustness by incorporating the countermeasures within its parameters during training. InI leverages a clone model during training as the surrogate adversary and estimates the adversary's optimization gradient and the expected gradient. With this estimation, the victim can learn to adjust its posterior probabilities to maximize the directional divergence between the two gradients. Therefore, we can isolate the clone model's optimization gradient from the expected gradient. To ameliorate the \emph{trade-off between benign accuracy and stealing robustness}, different from previous studies that add perturbations over all predictions that harm the benign accuracy, we aim to train models that can learn to behave differently on benign and malicious queries. Following the assumption of previous works~\cite{kariyappa2020defending, kariyappa2021protecting} that the malicious query samples deviate from the task distribution, InI trains a victim that behaves normally on the benign task yet produces inductive outputs on the malicious samples. Specifically, we introduce an out-of-distribution (OOD) dataset during victim training and minimize the adversary's information gain on it. As a result, during inference, the adversary is induced to extract little useful knowledge from the victim model using the stealing query with OOD examples. In addition, our method can be integrated with existing methods to better obtain defensive performance.

In summary, our \textbf{main contributions} are three-fold:
\begin{itemize}
    \item We propose a novel and effective defensive training framework against model stealing attacks called InI to achieve robustness during training, which provides a new perspective of model stealing defenses. 
    \item For computational overheads, InI incorporate the gradient isolation countermeasure within the victim's parameters; for unfavorable trade-offs, InI produces distinct outputs and induce the adversary to acquire minimal knowledge from malicious queries.
    \item Extensive experiments have been conducted on multiple datasets which demonstrate the state-of-the-art robustness and speed over other baselines. Moreover, InI shows flexible compatibility with existing methods for better defense.
\end{itemize}

\section{Related work}

\subsection{Model Stealing Attacks}
Model stealing attack, also referred to as model extraction, aims at inferring hyper-parameters~\cite{wang2018stealing, oh2019towards}, extracting model parameters~\cite{lowd2005adversarial, tramer2016stealing, milli2019model}, or copying functionalities of a certain machine learning model. Our work focuses on stealing the classification accuracy of the model, which is the most prevailing and universal stealing attack in deep learning. Tram\`er \textit{et al.}~\cite{tramer2016stealing} proposed the concept of model stealing that attackers could ``steal'' the property of a machine learning model by queries without prior knowledge of the victim model. Generally speaking, there are many properties that can be stolen, \textit{e.g.}, model parameters, training data, or functionality. Papernot \textit{et al.}~\cite{papernot2017practical} proposed a partial data approach named Jacobian-Based Dataset Augmentation (JBDA), which generates synthetic data by adding small perturbations on a small set of in-distribution samples. Orekondy \textit{et al.}~\cite{orekondy2019knockoff} proposed KnockoffNets, employing samples from a surrogate dataset as query inputs of the victim model. The stealing performance of partial data and surrogate data approaches would degrade when the available data are different from the original training set. In recent years, some data-free stealing methods have been proposed. Kariyappa \textit{et al.}~\cite{kariyappa2021maze} and Truong \textit{et al.}~\cite{truong2021data} are motivated by the framework of data-free knowledge distillation~\cite{hinton2015distilling, micaelli2019zero} and proposed data-free model stealing methods, where they use zeroth-order gradient estimation to calculate the victim gradient in black-box settings. Moreover, Sanyal \textit{et al.}~\cite{sanyal2022towards} proposed a model stealing attack in the hard-label setting, They utilize some unrelated proxy data to get a pre-trained data generator, while the stealing process is data-free.

\subsection{Model Stealing Defenses}

Currently, most model stealing defenses tend to add perturbations to the model outputs, thus disturbing the optimization of the adversary. Lee \textit{et al.}~\cite{lee2019defending} proposed an accuracy-preserving defense against model stealing attacks by adding deceptive perturbations to the model outputs while preserving its top-1 label, while it yields to the hard-label stealing. Other defenses like Maximizing Angular Deviation (MAD)~\cite{orekondy2019prediction} perturbs the model outputs with controllable intensity, defending against model stealing attacks at the expense of benign accuracy. Another approach of defense takes advantage of the data limitation of adversaries, making the victim model  produce dissimilar output between in-distribution inputs and out-of-distribution inputs. Kariyappa \textit{et al.}~\cite{kariyappa2020defending} proposed Adaptive Misinformation (AM) defense that detects the OOD inputs and misleads adversaries with modified outputs. Kariyappa \textit{et al.}~\cite{kariyappa2021protecting} then proposed Ensemble of Diverse Models (EDM) defense, which introduces randomness into the model output by using an ensemble of diverse models. Models in the ensemble are trained to perform diversely for OOD inputs, making the functionality of the model hard to be stolen. In addition, there are other types of countermeasures towards model stealing, such as digital watermarking~\cite{jia2021entangled, li2022defending, charette2022cosine}, which inject an extractable watermark into the victim model and can distinguish whether a model is from stealing.

Existing defensive approaches take advantage of some limitations of model stealing attacks to mitigate the knowledge leakage, while they are suffering from high computational costs and unfavorable trade-offs. In this paper, we are devoted to defending against stealing attacks by incorporating the countermeasure with the victim's parameters, inherently enhancing the robustness against model stealing attacks.

\section{Threat Model}

\subsection{Attack Objective}
In this paper, we mainly discuss the functionality stealing towards DNNs on image classification tasks. Specifically, an adversary aims at stealing the functionality of a victim model $\mathcal{V}$ by training a clone model $\mathcal{C}$. These attacks usually follow the framework of knowledge distillation~\cite{hinton2015distilling}, where the victim model $\mathcal{V}$ plays the role of ``teacher'', and the knowledge of the teacher is distilled into the ``student'' model $\mathcal{C}$. The objective of the adversary is to maximize the classification accuracy of clone model $Acc(\mathcal{C}(x;\bm{\theta}_\mathcal{C}), y)$ on the victim's target distribution $\mathcal{D}_{tar}$. Define $\bm{\theta}_\mathcal{V}$ to be the parameter of $\mathcal{V}$ and $\bm{\theta}_\mathcal{C}$ to be the parameter of $\mathcal{C}$, and the adversary's goal could be formulated as Eqn.~\ref{eqn:atk_goal}.

\begin{equation}
    \label{eqn:atk_goal}
    \max_{\bm{\theta}_\mathcal{C}}\mathbb{E}_{(\bm{x},\bm{y})\sim \mathcal{D}_{tar}}[Acc(\mathcal{C}(\bm{x}; \bm{\theta}_\mathcal{C}), \bm{y})]
\end{equation}

In most real-world settings, the adversary has no knowledge about the victim's structure, parameters, or training set. The only interaction between the adversary and the victim is the \textit{black-box query process}: the adversary inputs an image $x$ and the victim returns a softmax probability or logits. Though the original training set is unavailable, the adversary can use synthetic data or surrogate data to query the victim model. For example, JBDA~\cite{papernot2017practical} synthesizes data from a small part of in-distribution samples, and KnockoffNets~\cite{orekondy2019knockoff} uses surrogate datasets to query the victim model. Therefore, the adversary's learning objective is a surrogate goal based on the distribution of the query dataset $\mathcal{D}_{que}$, which can be formulated as follows:

\begin{equation}
    \label{eqn:atk_obj}
    \min_{\bm{\theta}_\mathcal{C}}\mathbb{E}_{\bm{x}\sim \mathcal{D}_{que}}[d(\mathcal{C}(\bm{x}; \bm{\theta}_\mathcal{C}), \mathcal{V}(\bm{x}; \bm{\theta}_\mathcal{V}))]
\end{equation}

In the ideal scenario, if $\mathcal{D}_{tar}$ and $\mathcal{D}_{que}$ are close enough and the query budget is sufficient, the model stealing is generally inevitable since the victim $\mathcal{V}$ must guarantee the performance for benign users on the target distribution. However, in practice, $\mathcal{D}_{tar}$ and $\mathcal{D}_{que}$ are dissimilar due to the knowledge limitation of the adversary, and the query budget is limited by the adversary's financial cost. It is these limitations that support existing defensive methods.

\subsection{Defense Objective} 
In defenses against model stealing, the defender aims at preventing the functionality of the victim model from being stolen with an acceptable impact on its benign accuracy. To be more practical, the accuracy degradation should be constrained with a minimum threshold $T$. The objective of the defender is to minimize the classification accuracy of the clone model $Acc(\mathcal{C}(x;\bm{\theta}_\mathcal{C}), y)$ on victim's target distribution $\mathcal{D}_{tar}$, which could be formulated as Eqn.~\ref{eqn:def_goal}.

\begin{equation}
\begin{aligned}
    \label{eqn:def_goal}
    \min_{\bm{\theta}_\mathcal{V}} \ & \mathbb{E}_{(\bm{x},\bm{y})\sim \mathcal{D}_{tar}}[Acc(\mathcal{C}(\bm{x}; \bm{\theta}_\mathcal{C}), \bm{y})], \\
    \text{s.t.} \ \  & \mathbb{E}_{(\bm{x},\bm{y})\sim \mathcal{D}_{tar}}[Acc(\mathcal{V}(\bm{x}; \bm{\theta}_\mathcal{V}), \bm{y})] \geq T
\end{aligned}
\end{equation}

Considering the limitations of the adversary, existing defense methods either add adaptive perturbations to multiply the adversary's query cost, or differentiate the victim's behavior on ID and OOD data to deceive the adversary. In this paper, we propose a defensive method against model stealing attacks, which gets rid of the extra computational costs and ameliorates the trade-off between benign accuracy and stealing resistance.

\section{Methodology}

\begin{figure*}[t]
\begin{center}
\includegraphics[width=0.9\linewidth]{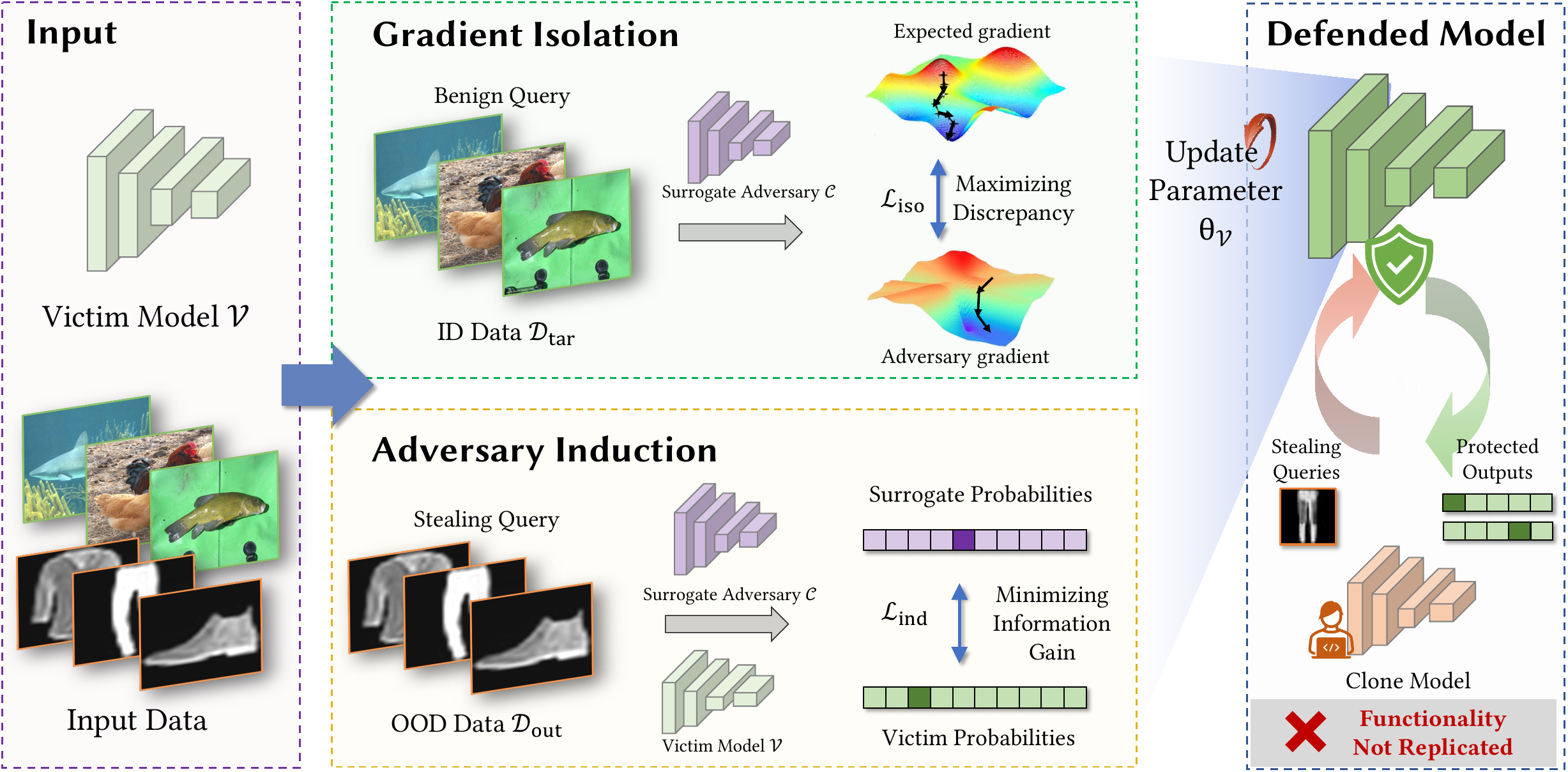}
\end{center}
\caption{The overall framework of InI. InI isolates the adversary's gradients from the expected gradients during training to obtain faster inference speed, and induce the adversary to leak knowledge as little as possible.} 
\label{fig:framework}
\end{figure*}

To build a defensive method with low computational costs and high trade-offs against model stealing attacks, we propose InI, a novel and effective defensive training framework. In this section, we first illustrate the training-time gradient isolation methodology that gets rid of auxiliary inference-time modules, and then elaborate on the adversary induction approach that reduces the knowledge leakage. Finally, we explain our overall training framework. 

\subsection{Gradient Isolation}

Existing defensive methods are suffering from extra computational costs during inference since they often employ auxiliary inference-time modules. Studies have revealed that DNNs are heavily over-parameterized~\cite{zhang2017understanding} and can be trained to fit and generalize across diverse data distributions, \emph{e.g.,} adversarial examples for adversarially-trained models~\cite{goodfellow2014explaining, madry2018towards, zhang2021interpreting, liu2021training, liu2023exploring, guo2023towards, tang2021robustart} and real-world disturbance for reinforcement agents~\cite{pinto2017robust, guo2022towards, li2023byzantine, li2023attacking}. Inspired by them, we propose a defensive training framework to directly train a robust model, so that the model can generate deceptive outputs towards stealing attack queries without extra modules.

Generally, as Eqn.~\ref{eqn:atk_goal} shows, the adversary's \textit{expected goal} is to minimize the disagreement with the ground truth on the target distribution. This objective is not directly related to the victim's parameter $\bm{\theta}_\mathcal{V}$, but the adversary must extract knowledge from the victim. As a consequence, the \textit{real objective} of the adversary is illustrated in Eqn.~\ref{eqn:atk_obj}. There exists a gap between the expected goal and the real objective, and the defense can be achieved by isolating the adversary's real objective from the expected goal. As the adversary usually updates its parameters by gradient descent, we propose gradient isolation to isolate the real gradient from the expected gradient, thereby incorporating the robustness within the victim's parameters to mislead the adversary.

To achieve gradient isolation, we need to estimate the above two gradient terms during training. Therefore, we introduce a surrogate white-box clone model $\mathcal{C}$ into the victim's training process to represent the stealing role of the adversary. To isolate the update gradient from the expected gradient, we maximize the directional divergence between them. Specifically, for a certain batch of data $\bm{x}$, assuming the target of the adversary is $\bm{y}$, the update gradient can be written as:

\begin{equation}
\begin{aligned}
    \nabla_{\bm{\theta}_\mathcal{C}} CE(\mathcal{C}(\bm{x}; \bm{\theta}_\mathcal{C}), \bm{y}) & = - \nabla_{\bm{\theta}_\mathcal{C}} \sum_i {y_i\log \mathcal{C}(\bm{x}; \bm{\theta}_\mathcal{C})_i} \\
    & = - \bm{y}^T \bm{G},
\end{aligned}
\end{equation}

\noindent where $CE(\cdot, \cdot)$ is the soft cross-entropy loss commonly used in model stealing attacks~\cite{papernot2017practical, orekondy2019knockoff}, and $\bm{G} = \nabla_{\bm{\theta}_\mathcal{C}} \log \mathcal{C}(\bm{x}; \bm{\theta}_\mathcal{C})$ is a Jacobian matrix. When $y$ comes from the ground truth, it represents the correct optimization direction; when $y$ comes from the victim model (denoted by $\tilde{\bm{y}}$), it represents the actual optimization direction during stealing. 

The goal of gradient isolation is to maximize the directional divergence of these gradients, which can be quantified by the cosine similarity. Therefore, the objective of gradient isolation can be written as follows:

\begin{equation}
\label{eqn:isolation}
    \mathcal{L}_{iso} = \mathbb{E}_{(\bm{x},\bm{y})\sim \mathcal{D}_{tar}} [CS(\tilde{\bm{y}}^T \bm{G} , \bm{y}^T \bm{G})],
\end{equation}

\noindent where $CS(\cdot, \cdot)$ represents the cosine similarity. $\tilde{\bm{y}}^T \bm{G}$ and $\bm{y}^T \bm{G}$ can be calculated by the backward propagation. During the victim training, InI minimizes $\mathcal{L}_{iso}$ via optimization to perform gradient isolation, enhancing the victim's robustness against model stealing attacks.

\subsection{Adversary Induction}

Some existing defenses add perturbations to the output over all samples, which harm the benign accuracy and cause low trade-offs. To improve the trade-off between clean accuracy and stealing robustness, we further design an adversary induction approach to train victim models. Thus, the victim model will behave normally for benign users but produce inductive outputs that induce the adversary to optimize without learning too much useful knowledge.

Successfully inducting the adversary stands on the assumption that benign and malicious queries can be distinguished through the distribution \cite{kariyappa2020defending, kariyappa2021protecting}. Practically, the adversary has limited knowledge of the distribution of the victim's training set $\mathcal{D}_{tar}$, and consequently uses a surrogate dataset $\mathcal{D}_{que}$ to query and steal the victim model. Following this assumption, we regard the query samples of the adversary as out-of-distribution and apply an OOD dataset $\mathcal{D}_{out}$ during defense to substitute them. The over-parameterization property of DNNs ensures the generalization capability across diverse distributions and thus enables the acquisition of the victim that exhibits divergent behavior on benign in-distribution (ID) queries and malicious out-of-distribution (OOD) queries. Thus, our InI aims to guarantee the victim's benign performance on ID queries, while inducing the adversary to attain little knowledge with uninformative outputs on OOD queries.

In particular, on ID samples, we should guarantee the victim's benign performance. This can be achieved by applying a cross-entropy loss to train the classification model, which is shown as follows:

\begin{equation}
\label{eqn:benign}
    \mathcal{L}_{ben} = \mathbb{E}_{(\bm{x},\bm{y})\sim \mathcal{D}_{tar}} [CE(\mathcal{V}(\bm{x}; \bm{\theta}_\mathcal{V}), \bm{y})].
\end{equation}

On OOD samples, we should induce the adversary to acquire minimal knowledge. Intuitively, the adversary induction can be realized by reducing the adversary's information gain on OOD queries. The information gain can be quantified by the KL divergence between the output probabilities of the clone and the victim model, which can be formulated as below:

\begin{equation}
\label{eqn:infogain}
    \mathcal{L}_{ig} = \mathbb{E}_{\bm{x}\sim \mathcal{D}_{out}} [KL(\mathcal{C}(\bm{x}; \bm{\theta}_\mathcal{C}), \mathcal{V}(\bm{x}; \bm{\theta}_\mathcal{V}))].
\end{equation}

Noted that $\mathcal{L}_{ig}$ is the optimization objective of the adversary during model stealing. Therefore, when $\mathcal{L}_{ig}$ is minimized before the stealing process, the adversary can only gain little information via optimization. Thus, the adversary would only extract little knowledge from the victim model. Since the adversary often uses gradient descent to update their parameters and attain knowledge, minimizing the first-order approximation of the KL divergence can also help to reduce the information gain. Consequently, we can minimizing the norm of $\nabla_{\bm{\theta}_\mathcal{C}} \mathcal{L}_{ig}$, \textit{i.e.}, the gradient of the information gain. In summary, the objective of adversary induction can be formulated as:

\begin{equation}
\label{eqn:induction}
    \mathcal{L}_{ind} = \mathcal{L}_{ig} + \beta ||\nabla_{\bm{\theta}_\mathcal{C}} \mathcal{L}_{ig}||.
\end{equation}

$\mathcal{L}_{ind}$ is a function related to $\bm{\theta}_\mathcal{V}$, which can be integrated with our training framework. During the victim's training, we apply a surrogate white-box clone model $\mathcal{C}$ and an OOD dataset to estimate the adversary's information gain and minimize $\mathcal{L}_{ind}$ by updating the victim's parameter $\bm{\theta}_\mathcal{V}$, thus reducing the knowledge leakage from the victim. The OOD dataset used in training comes from another classification task and is different from the query dataset in stealing attacks.

Though previous defenses~\cite{kariyappa2020defending, kariyappa2021protecting} utilize OOD datasets to train the victim, they only constrain the victim to produce meaningless or diverse outputs and did not take the adversary's role into consideration. In contrast to them, InI produces inductive outputs that trigger the adversary to learn less, reducing the knowledge leakage from the victim.

\subsection{Overall Framework}

Figure~\ref{fig:framework} illustrates our overall framework. To train a robust victim $\mathcal{V}$, we introduce a surrogate clone model $\mathcal{C}$ into the training process. During training, the victim has white-box access to the surrogate clone model. To jettison the auxiliary modules for low computational costs, InI isolates the adversary's optimization gradient from the expected gradient during training via $\mathcal{L}_{iso}$ in Eqn.~\ref{eqn:isolation}. To further improve the trade-off between benign performance and stealing robustness, InI produces uninformative outputs on malicious queries through $\mathcal{L}_{iso}$ in Eqn.~\ref{eqn:induction} to induce the adversary to acquire less useful knowledge. By incorporating the robustness within the victim's parameters, the victim will acquire how to resist the model stealing attacks in a favorable trade-off without any auxiliary modules during inference.

During training, all objectives are simultaneously calculated and updated. We use some hyper-parameters $\gamma_1$, $\gamma_2$ to control the trade-off between each loss, and the total loss can be written as:

\begin{equation}
    \mathcal{L} = \mathcal{L}_{ben} + \gamma_1 \mathcal{L}_{iso} + \gamma_2 \mathcal{L}_{ind}
\end{equation}

\begin{figure}[t]
\begin{center}
    \includegraphics[width=0.9\linewidth]{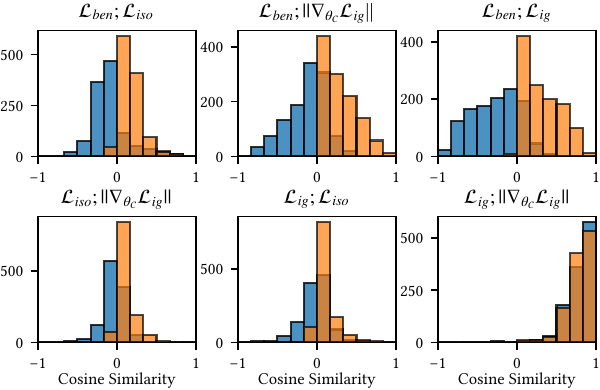}
\end{center}
\caption{The cosine similarities of objective pairs during training. We choose the gradients of $\mathcal{L}_{ben}$, $\mathcal{L}_{ind}$,  $\nabla_{\bm{\theta}_\mathcal{C}} \mathcal{L}_{ind}$ and $\mathcal{L}_{iso}$ at the first 3 epochs, and show the histogram of their cosine similarities. \textcolor{blue}{Blue} bars indicates cosine similarities before the gradient surgery, and \textcolor{orange}{orange} bars indicates those after the gradient surgery.}
\label{fig:grad}
\end{figure}

However, during training, there may exist some conflicts between the optimization directions among the above losses. As the blue bars shown in Figure~\ref{fig:grad}, we extract the gradients at the first 3 epochs of training and calculate their cosine similarities, where we observe that the cosine similarities between different objectives exist conflicts. To mitigate these conflicts, we leverage PCGrad~\cite{yu2020gradient} to deal with the gradient conflicts. Before the gradient descent update, PCGrad tries to find the conflict among these gradients and projects one gradient to the orthogonal direction of the others as follows:

\begin{equation}
    \bm{g}^{PC}_{i} = \bm{g}^{PC}_{i} - \frac{\bm{g}^{PC}_{i} \cdot \bm{g}_{j}}{||\bm{g}_{j}||^2} \bm{g}_{j}.
\end{equation}

After PCGrad stabilization, the conflicts are better mitigated and the optimization process is improved (see the orange bars in Figure~\ref{fig:grad}). The overall pseudo-algorithm of our InI can be found in Supplementary Material. 

\section{Experiments}

In this section, we first elaborate on the experimental settings; then, we illustrate the defense performance and inference speed analysis on image classification tasks; we finally provide ablation studies. \cready{More results such as feature visualization analysis of our defense are provided in Supplementary Materials.}

\subsection{Experimental Setup}

In this part, we elaborate on our experimental settings about datasets, model architectures, defenses, attacks, and evaluation metrics.

\textbf{Datasets and architectures.} We evaluate our proposed InI on the most commonly-adopted image classification datasets for model stealing including MNIST~\cite{lecun1998gradient}, FashionMNIST~\cite{xiao2017fashion}, CIFAR-10~\cite{krizhevsky2009learning}, and CIFAR-100~\cite{krizhevsky2009learning}. We choose ResNet-18~\cite{he2016deep} as the backbone of all victim models. We also evaluate results on VGG networks~\cite{simonyan2014very} which show similar observations (\emph{c.f.} Supplementary Materials). 

\textbf{Implementation details.} For the training of InI, we use an SGD optimizer with momentum 0.5 and a weight decay of $1 \times 10^{-3}$. For MNIST and FashionMNIST datasets, we train 50 epochs with a learning rate annealing of 0.1 every 20 epochs, and for CIFAR-10 and CIFAR-100 datasets, we train 150 epochs with a learning rate annealing of 0.1 every 50 epochs. The initial learning rate is 0.1.

\textbf{Defenses.} To demonstrate the effectiveness of InI, we compare our method with the commonly-adopted defensive approaches: MAD~\cite{orekondy2019prediction}, AM~\cite{kariyappa2020defending}, EDM~\cite{kariyappa2021protecting}. We also report the results of an undefended model denoted by ``Vanilla''. We referred to the official implementation of these methods. The batch size of all defensive methods is 128. For the auxiliary OOD datasets used by AM, EDM, and InI, we choose KMNIST~\cite{clanuwat2018deep} for MNIST and FashionMNIST, and choose TinyImageNet~\cite{le2015tiny} for CIFAR-10 and CIFAR-100. For the hash dataset used by EDM, we choose KMNIST for MNIST and FashionMNIST, and use SVHN~\cite{netzer2011reading} for CIFAR-10 and CIFAR-100. 

\textbf{Attacks.} Following the previous works\cite{orekondy2019prediction, kariyappa2020defending, kariyappa2021protecting}, to evaluate the performance of InI against model stealing, we use the commonly-used stealing attacks including KnockoffNets~\cite{orekondy2019knockoff} and JBDA~\cite{papernot2017practical}, and evaluate the integration of InI with other defensive methods. For each attack, we conduct soft-label and hard-label attacks, which means the adversary learns according to the victim's output probability and the top-1 label, respectively. The detailed settings of attacks are listed as follows:

\begin{itemize}
    \item \textit{KnockoffNets}: The budget of attack is 50000. As for the surrogate datasets, we use EMNISTLetters~\cite{cohen2017emnist}, EMNIST~\cite{cohen2017emnist}, CIFAR-100, CIFAR-10 as the surrogate dataset for MNIST, FashionMNIST, CIFAR-10, CIFAR-100.
    \item \textit{JBDA}: We choose 150 images from the victim's training set as the seed samples. We use 6 rounds of augmentation and a noise rate of 0.1 to synthesize the query data. The clone model is trained for 10 epochs every augmentation round.
\end{itemize}

\textbf{Evaluation metrics.} Following \cite{kariyappa2021protecting}, we evaluate the performance of defenses by comparing the \emph{clone accuracy} achieved by attacks on the victim's test set. To take the victim's benign accuracy into consideration, we further compare the \emph{relative performance} of the defenses, which is the ratio of the adversary's clone accuracy and the victim's benign accuracy. For methods that have mutable parameters during inference (\textit{i.e.}, MAD and AM), we follow the setting in~\cite{kariyappa2021protecting} and adjust the parameters of the defense to have similar benign accuracy on the test set. The benign accuracy of these defenses on classification tasks is shown in Table~\ref{tab:benign}. \emph{For all the above metrics, the lower the better defenses.}

\begin{table}
    \caption{Benign accuracy of each defense on different image classification datasets.}
    \label{tab:benign}
    \begin{tabular}{@{}ccccc@{}}
    \toprule
    \multirow{2}{*}{Dataset} & \multicolumn{4}{c}{Benign Accuracy}       \\ \cmidrule(l){2-5} 
                             & MNIST & FashionMNIST & CIFAR-10 & CIFAR-100 \\ \midrule
    Vanilla                       & 99.46 & 93.89        & 94.71   & 76.63    \\
    MAD                      & 99.46 & 93.87        & 94.31   & 75.44    \\
    AM                       & 99.41 & 93.67        & 94.30   & 75.00    \\
    EDM                      & 99.43 & 93.70        & 94.35   & 75.38    \\
    InI (\textbf{Ours})  & 99.40 & 93.36        & 94.32   & 75.50    \\ \bottomrule
    \end{tabular}
\end{table}

\begin{table*}
    \caption{Experimental results for KnockoffNets attack. We report the clone accuracy and the relative performance on the target test set. Lower clone accuracy/relative performance indicates better defense performance. ``InI + MAD'' and ``InI + AM'' indicate the integration of our InI with other defenses.}
    \label{tab:knockoff}
    \begin{tabular}{@{}ccccccccc@{}}
    \toprule
    \multirow{2}{*}{Defense} & \multicolumn{2}{c}{MNIST}                     & \multicolumn{2}{c}{FashionMNIST}              & \multicolumn{2}{c}{CIFAR-10}                   & \multicolumn{2}{c}{CIFAR-100}                 \\ \cmidrule(l){2-9} 
                             & soft-label            & hard-label            & soft-label            & hard-label            & soft-label            & hard-label            & soft-label           & hard-label            \\ \midrule
    Vanilla                  & 99.39(1.00×)          & 98.84(0.99×)          & 71.58(0.76×)          & 57.95(0.62×)          & 79.55(0.84×)          & 69.60(0.73×)          & 50.89(0.66×)         & 27.44(0.36×)          \\
    MAD                      & 99.31(1.00×)          & 99.05(1.00×)          & 68.84(0.73×)          & 44.61(0.48×)          & 70.31(0.75×)          & 65.07(0.69×)          & 37.36(0.50×)         & 18.58(0.25×) \\
    AM                       & 98.58(0.99×)          & 97.14(0.98×)          & 20.77(0.22×)          & 14.23(0.15×)          & 75.32(0.80×)          & 63.08(0.67×)          & 24.07(0.32×)         & \textbf{15.99(0.21×)}          \\
    EDM                      & 98.90(0.99×)          & 97.44(0.98×)          & 21.42(0.23×)          & 15.90(0.17×)          & 72.30(0.77×)          & 62.31(0.66×)          & 43.78(0.58×)         & 20.52(0.27×)          \\
    InI (\textbf{Ours})   & \textbf{89.02(0.90×)} & \textbf{95.90(0.96×)} & \textbf{20.12(0.22×)} & \textbf{10.82(0.12×)} & \textbf{69.54(0.74×)} & \textbf{60.33(0.64×)} & \textbf{9.71(0.13×)} & 22.01(0.29×)          \\ \midrule \midrule
    InI + MAD           & \textbf{88.09(0.89×)} & \textbf{92.50(0.93×)} & 20.18(0.22×) & 10.77(0.12×) & 67.45(0.72×) & 60.25(0.64×) & 9.37(0.12×) & 13.06(0.17×)          \\ 
    InI + AM            & 88.22(0.89×) & 94.12(0.95×) & \textbf{15.01(0.16×)} & \textbf{10.27(0.11×)} & \textbf{65.80(0.70×)} & \textbf{58.35(0.62×)} & \textbf{9.36(0.13×)} & \textbf{12.47(0.17×)}   \\ \bottomrule
    \end{tabular}
\end{table*}

\begin{table*}
    \caption{Experimental results for JBDA attack. We report the clone accuracy and the relative performance on the target test set. Lower clone accuracy/relative performance indicates better defense performance. ``InI + MAD'' and ``InI + AM'' indicate the integration of our InI with other defenses.}
    \label{tab:jbda}
    \begin{tabular}{@{}ccccccccc@{}}
    \toprule
    \multirow{2}{*}{Defense} & \multicolumn{2}{c}{MNIST}                     & \multicolumn{2}{c}{FashionMNIST}              & \multicolumn{2}{c}{CIFAR-10}                   & \multicolumn{2}{c}{CIFAR-100}                \\ \cmidrule(l){2-9} 
                             & soft-label            & hard-label            & soft-label            & hard-label            & soft-label            & hard-label            & soft-label           & hard-label           \\ \midrule
    Vanilla                  & 73.00(0.73×)          & 72.77(0.73×)          & 71.09(0.76×)          & 67.80(0.72×)          & 26.19(0.28×)          & 25.59(0.27×)          & 4.82(0.06×)          & 4.09(0.05×)          \\
    MAD                      & 61.69(0.62×)          & 72.81(0.73×)          & \textbf{57.70(0.61×)} & 66.46(0.71×)          & \textbf{18.73(0.20×)} & 24.89(0.26×)          & \textbf{2.44(0.03×)} & 3.90(0.05×)          \\
    AM                       & 81.23(0.82×)          & 73.17(0.74×)          & 67.73(0.72×)          & 66.28(0.71×)          & 24.33(0.26×)          & 25.12(0.27×)          & 4.36(0.06×)          & 3.29(0.04×)          \\
    EDM                      & 79.34(0.80×)          & 78.72(0.79×)          & 70.08(0.75×)          & 68.86(0.73×)          & 25.86(0.27×)          & 25.71(0.27×)          & 3.35(0.04×)          & 3.04(0.04×)          \\
    InI (\textbf{Ours})  & \textbf{57.94(0.58×)} & \textbf{66.19(0.67×)} & 70.81(0.76×)          & \textbf{64.99(0.70×)} & 24.16(0.26×)          & \textbf{24.14(0.26×)} & 3.81(0.05×)          & \textbf{2.61(0.03×)} \\ \midrule \midrule
    InI + MAD           & \textbf{55.99(0.56×)} & 66.09(0.66×) & \textbf{65.63(0.70×)} & 64.61(0.69×) & 23.04(0.24×) & 24.04(0.25×) & 3.51(0.05×) & 2.61(0.03×)          \\ 
    InI + AM            & 56.42(0.57×) & \textbf{62.35(0.63×)} & 68.14(0.73×) & \textbf{63.72(0.68×)} & \textbf{22.68(0.24×)} & \textbf{22.43(0.24×)} & \textbf{3.33(0.04×)} & \textbf{2.45(0.03×)}   \\ \bottomrule
    \end{tabular}
\end{table*}

\subsection{Defense Results on Stealing Attacks}

In this part, we compare the performance of our InI with other model stealing defenses. We report the clone accuracy and the relative performance of existing defenses and InI in Tables~\ref{tab:knockoff} and \ref{tab:jbda}. To further improve our defensive performance, we integrate the model trained by InI with MAD and AM and evaluate the performance. AM needs to jointly train the victim, yet we do not retrain our model in their framework but load the parameters from InI to replace the victim's parameter for simplicity.

\textbf{KnockoffNets attacks.} Table~\ref{tab:knockoff} presents the defense results against KnockoffNets attacks. We can draw some observations listed below:
\begin{itemize}
    \item By inducing and isolating the adversary, InI achieves the best defense performance with similar benign accuracy on most of the results, which demonstrates our better trade-offs. 
    \item Noted that, on CIFAR-100 dataset, InI achieves an extraordinary defense performance against the soft-label attack, but behaves poorly against the hard-label attack. 
\end{itemize}

\textbf{JBDA attacks.} Table~\ref{tab:jbda} shows the defense performance against JBDA attacks. Different from KnockoffNets which applys surrogate datasets, JBDA uses a set of seed examples from the victim's training set, and thus the results differ from KnockoffNets. From the results, we can observe that:

\begin{itemize}
    \item The OOD-based defenses (AM, EDM, and InI) behave poorly on the soft-label JBDA. This mainly results from that the samples of JBDA come from the seed examples in the training set and are close to the target distribution.
    \item Instead, MAD, the perturbation-based method, could achieve the best performance on the soft-label attack on FashionMNIST, CIFAR-10, and CIFAR-100 datasets. However, its performance rapidly drops on the hard-label attacks, as it hardly modifies the top-1 label of the output.
    \item JBDA achieves a good stealing performance on simpler datasets like MNIST and FashionMNIST, but cannot behave well on more complex datasets like CIFAR-10 and CIFAR-100. The JBDA stealing results on CIFAR-100 (around 0.05× on all defenses) is generally unusable.
\end{itemize}

\textbf{Integration with other defenses.} We also provide the experimental results of the integration of our InI with MAD and AM in Table~\ref{tab:knockoff} and \ref{tab:jbda}. We can draw some observations that:
\begin{itemize}
    \item The integration of our InI with MAD and AM can further improve the trade-off, as they achieve further defense performance with invariant benign accuracy than the original MAD and AM against all the attacks. Additionally, the integration of our InI with AM mostly achieves more robustness than that with MAD.
    \item On soft-label JBDA attacks, though the integration achieves improvement in Vanilla and InI, it still cannot surpass the performance of MAD in similar benign accuracy. We attribute this phenomenon to the robustness trade-off between soft-label and hard-label attacks, as InI has traded its benign accuracy off the robustness against hard-label attacks during training, and similar robustness can only be achieved at the cost of more benign performance.
\end{itemize}

\subsection{Inference Speed Analysis} 

In this section, we provide the inference speed analysis of each defensive method and evaluate our analysis through experiments. In practice, the model for MLaaS would only train once but would receive millions of inference queries from users, which the service provider would charge for. Therefore, boosting the inference speed has a significant impact on the practical use of defensive methods. \cready{Evaluations and discussions about training-time speed are provided in Supplementary Materials.}

We first provide some analyses of the inference process of each method. As shown in Table~\ref{tab:speed}, our InI can achieve the fastest inference speed among all defensive methods. On the contrary, existing methods employ auxiliary modules during inferences, which would introduce extra computational operations, or even harm the DNN's parallel capability. Detailed analyses are given below.

\textbf{Time cost of Vanilla and InI.} We define the time cost of a forward operation as $M_f$ for a single query. The time cost of a model without defense should be $M_f$, and the same as InI since InI employs no extra modules. 

\textbf{Time cost of MAD.} MAD computes the Jacobian matrix $G=\nabla \log f(x, \bm{\theta})$ to maximize the angular deviation between the perturbed gradient and the original gradient. The official code needs $C$ backward passes which costs $CM_b$ ($C$ is the number of classes of the target classification task) to calculate the Jacobian matrix. After getting $G$, MAD needs to heuristically search a perturbed probability $y^*$, which costs time of $S$. When the victim receives a batch of data with a mini-batch size of $B$, the backward pass and heuristic search would cost $BCM_b$ and $BS$, as these operations cannot perform in parallel.

\textbf{Time cost of AM.} AM employs an adaptive misinformation module to perturb the victim's output probability. The adaptive misinformation is generated by a DNN with the same architecture as the victim's backbone. As a consequence, the time cost of AM mainly comes from the forward passes of the victim's backbone and the misinformation model, which cost around $2M_f$ in sum.

\begin{table}[t]
    \caption{Analysis for the inference time of each defense.}
    \setlength{\tabcolsep}{14pt}
    \label{tab:speed}
    \begin{tabular}{@{}ccc@{}}
    \toprule
              & Single Query  & Batch Query           \\ \midrule
    Vanilla/InI   & $M_f$             & $M_f$             \\
    MAD       & $M_f+CM_b+S$      & $M_f+B(CM_b+S)$   \\
    AM        & $2M_f$            & $2M_f$            \\
    EDM       & $M_f+M_h$         & $nM_f+M_h$        \\ \bottomrule
    \end{tabular}
\end{table}

\begin{figure}[t]
    \centering
    \includegraphics[width=0.9\linewidth]{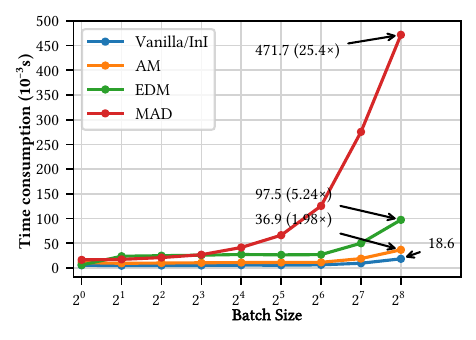}
    \caption{Empirical evaluation of inference speed performance of each defense approaches.}
    \label{fig:speed}
\end{figure}

\textbf{Time cost of EDM.} EDM jointly trains an ensemble of $n$ victim models and selects a result from them according to a hash function. The hash function of EDM is a DNN with a simpler architecture, which costs $M_h$ for a forward pass. For a single query, the time cost of EDM should be $M_f+M_h$. However, in the official implementation of EDM, the time cost increase to $nM_f+M_h$ when EDM receives a batched query. The reason lies in that the forward pass of a single model and a batch of data can run in parallel, but the forward pass of different models cannot run in parallel. Therefore, all models in the ensemble would be accessed, and the time cost would increase.

\textbf{Empirical evaluation.} We perform empirical experiments to evaluate the speed of each defense. The experiment is conducted on an RTX 3080 GPU and the architecture of the victim's backbone is ResNet-18. We randomly generate input images, execute the inference, and record the time cost. For fair comparisons, we conduct the process repeatedly for 2000 times and record the total time. As shown in Figure~\ref{fig:speed}, other defenses consume significantly more time during inference compared to our method (1.98$\times$-25.4$\times$).

\subsection{Ablation Studies}

\begin{table}[t]
    \centering
    \caption{Ablation studies on the contributions of gradient isolation and adversary induction. Results are shown in clone accuracy and relative performance (lower the better).}
    \label{tab:ablation}
    \small
    \tabcolsep=0.1cm
    \begin{tabular}{@{}ccccc@{}}
    \toprule
    \multirow{2}{*}{Defense} & \multicolumn{2}{c}{KnockoffNets} & \multicolumn{2}{c}{JBDA} \\ \cmidrule(l){2-5} 
                             & soft-label      & hard-label     & soft-label  & hard-label \\ \midrule
    Vanilla                  & 79.55(0.84×)  & 69.60(0.73×)   & 26.19(0.28×)  & 25.59(0.27×) \\
    $\mathcal{L}_{iso}$                & 75.21(0.79×)  & 67.48(0.71×)   & 25.30(0.27×)  & 25.37(0.27×)\\
    $\mathcal{L}_{ind}$                & 78.35(0.83×)   & 65.92(0.70×)  & 24.79(0.26×) & 24.76(0.26×) \\
    \cready{w/o $\nabla \mathcal{L}_{ig}$}      & 74.59(0.79×)   & 64.73(0.68×)  & 24.55(0.26×) & 25.15(0.27×) \\
    $\mathcal{L}_{iso}$+$\mathcal{L}_{ind}$ (InI) & 69.54(0.74×) & 60.33(0.64×)  & 24.16(0.26×)  & 24.14(0.26×)\\ \bottomrule
    \end{tabular}
\end{table}

We then conduct ablation studies to understand the contributions of gradient isolation and adversary induction. Specifically, we conduct experiments by training the victim with (1) no extra defense (denoted by ``Vanilla''); (2) only isolation loss $\mathcal{L}_{iso}$; (3) only induction loss $\mathcal{L}_{ind}$; \cready{(4) InI without $\nabla \mathcal{L}_{ig}$; and (5)} the full InI ($\mathcal{L}_{iso}$+$\mathcal{L}_{ind}$). We record the defense performance against KnockoffNets and JBDA on the CIFAR-10 dataset. As shown in Table~\ref{tab:ablation}, we can draw several observations: 

\begin{itemize}
    \item Model trained with $\mathcal{L}_{iso}$ has an apparent drop in stealing performance (\textit{e.g.}, 4.34 on soft-label KnockoffNets), which indicates that gradient isolation can bring robustness against model stealing. 
    \item Model trained with $\mathcal{L}_{ind}$ shows limited defenses against model stealing attacks, as it only induces the adversary to learn less instead of directly misleading the adversary.
    \item With the two methods combined, InI can achieve the best defense performance (69.54 / 60.33 on KnockoffNets and 24.16 / 24.14 on JBDA), implying that the cooperation of gradient isolation and adversary induction plays an important role during training.
\end{itemize}

\section{Conclusion}

The model stealing attack becomes a raising challenge to the privacy and intellectual property of machine learning models. Existing defensive methods are suffering from additional computational costs and unfavorable trade-offs, which impede their practical implementation. To cope with this concern, we propose a novel and efficient training framework named InI. InI embeds the countermeasures within the victim's parameters, isolating the adversary's gradient from the expected gradient to achieve robustness without incurring extra computational overheads. InI leverages the OOD assumption and induces the adversary to acquire minimal knowledge, thereby enhancing the trade-off. Through our evaluations, InI surpasses existing defensive methods in terms of speed and robustness, and the integration with prior defenses renders it more practical. We hope our proposed method could provide a new perspective of defense strategies against model stealing attacks.

\section*{Acknowledgement} 

This work was supported by the National Natural Science Foundation of China (62022009 and 62206009), the Fundamental Research Funds for the Central Universities, and the State Key Laboratory of Software Development Environment.

\normalsize
\bibliographystyle{ACM-Reference-Format}
\balance
\bibliography{sample-base}

\newpage
\appendix
\counterwithin{table}{section}
\counterwithin{figure}{section}

\section*{Appendix}

\section{Pseudo Code of InI}

\label{pseudocode}

\begin{algorithm}[h]
    \caption{InI: Defense against Model Stealing}
    \label{alg:ours}
    \begin{algorithmic}[1]
    \renewcommand{\algorithmicrequire}{\textbf{Input:}}
    \renewcommand{\algorithmicensure}{\textbf{Output:}}
    \REQUIRE Number of iterations $N$, induction coefficient $\beta_0, \beta_1$, isolation coefficient $\gamma$, training set $\mathcal{D}_{tar}$, surrogate query dataset $\mathcal{D}_{que}$
    \ENSURE The defended model $\mathcal{V}$
    \STATE Initialize $\bm{\theta}_\mathcal{V}, \bm{\theta}_\mathcal{C}$
    \FOR {iteration \textbf{in} $N$}
        \STATE Sample a mini-batch of data $(\bm{x}, \bm{y})$ from $\mathcal{D}_{tar}$
        \STATE Calculate benign utility loss $\mathcal{L}_{ben} = CE(\mathcal{V}(\bm{x}; \bm{\theta}_\mathcal{V}), \bm{y})$
        \STATE Calculate isolation loss $\mathcal{L}_{iso} = CS(\tilde{\bm{y}}^T \bm{G} , \bm{y}^T \bm{G})$
        \STATE Sample a mini-batch of data $x$ from $\mathcal{D}_{que}$
        \STATE Calculate information gain $\mathcal{L}_{ig} = d(\mathcal{C}(\bm{x}; \bm{\theta}_\mathcal{C}), \mathcal{V}(\bm{x}; \bm{\theta}_\mathcal{V}))$
        \STATE Get the gradient of information gain $\nabla_{\bm{\theta}_\mathcal{C}} \mathcal{L}_{ig}$ through backward propagation
        \STATE Mitigate the gradient conflict of $\mathcal{L}_{ben}$, $\mathcal{L}_{iso}$, $\mathcal{L}_{ig}$, $||\nabla_{\bm{\theta}_\mathcal{C}} \mathcal{L}_{ig}||$ using PCGrad
        \STATE Update $\bm{\theta}_\mathcal{V}$ with SGD optimizer
    \ENDFOR
    \end{algorithmic}
\end{algorithm}

\section{Training Time Evaluation}

Though InI makes great progress on inference time speed improvement, the operation of calculating gradients increases its training overheads. We evaluate the training time of each defense method on CIFAR-10, and record the time consumption of 10 epochs. The results are listed in Table~\ref{tab:train_time}. The training time is 12.15× as undefended training and 1.40× time consumption as EDM, as we compute different gradients to construct the isolation loss. However, we achieve most 25.4x speed improvement than baseline methods during inference, and for MLaaS models, the proportion of inference time is much more than training time. 

\begin{table}[!htbp]
\caption{The time consumption of 10 epochs training for each defense methods.}
\label{tab:train_time}
\small
\begin{tabular}{@{}ccccc@{}}
\toprule
     & ND/MAD & AM      & EDM     & InI(\textbf{Ours}) \\ \midrule
Time(s) & 155.7(1.00×) & 551.1(3.54×) & 1353.5(8.69×) & 1892.1(12.15×)    \\ \bottomrule
\end{tabular}
\end{table}

\section{Training the Surrogate Adversary}

In our main experiments of InI, the surrogate adversary $\mathcal{C}$ keeps untrained, which is similar to~\cite{orekondy2019prediction}. We also perform adversarial training for the surrogate adversary $\mathcal{C}$, where $\mathcal{C}$ and $\mathcal{V}$ have opposite objectives and the parameters $\theta_\mathcal{C}$ and $\theta_\mathcal{V}$ are alternately updated. The results are shown in Table~\ref{tab:adv_training}. We observe that adversarial training is very time-consuming and the defensive performance is sensitive to hyperparameters. We can draw from the results that adversarial training either has poor defensive performance or harms benign accuracy. The unstable performance of adversarial training makes it almost unusable.

\begin{table}[b]
\caption{The defense result when adversarially training the surrogate adversary. The steal accuracy is from KnockoffNets soft-label attack on CIFAR-10 datasets. ``Adv Training 1'' an ``Adv training 2'' refer to different hyperparameters.}
\label{tab:adv_training}
\begin{tabular}{@{}ccc@{}}
\toprule
Defense        & Clean accuracy & Steal accuracy \\ \midrule
Vanilla        & 94.71          & 79.55          \\
InI            & 94.32          & 69.54          \\
Adv Training 1 & 94.66          & 77.23          \\
Adv Training 2 & 64.95          & 40.40          \\ \bottomrule
\end{tabular}
\end{table}

\section{Feature Visualization} 

To better understand the effectiveness of our defense, we further conduct feature visualizations of ID and OOD samples for the victim models. Specifically, we pick 5,000 images from the test set of ID (MNIST) and OOD (EMNISTLetters) datasets and use the convolutional layers of victim models to extract the features, and finally plot them using t-SNE~\cite{van2008visualizing}.

\begin{figure}[!htbp]
\begin{center}
\subfigure[Vanilla]{
    \includegraphics[width=0.45\linewidth]{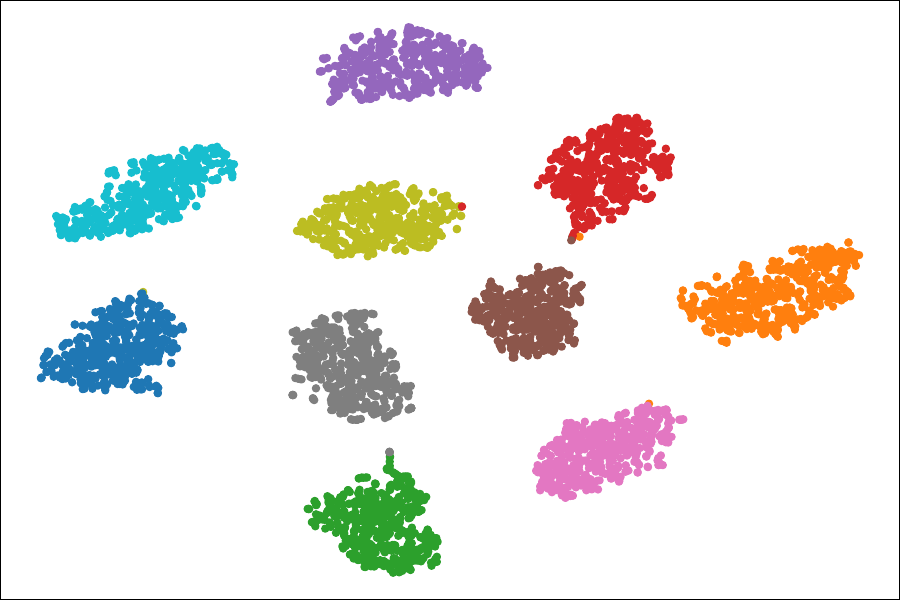}
    \includegraphics[width=0.45\linewidth]{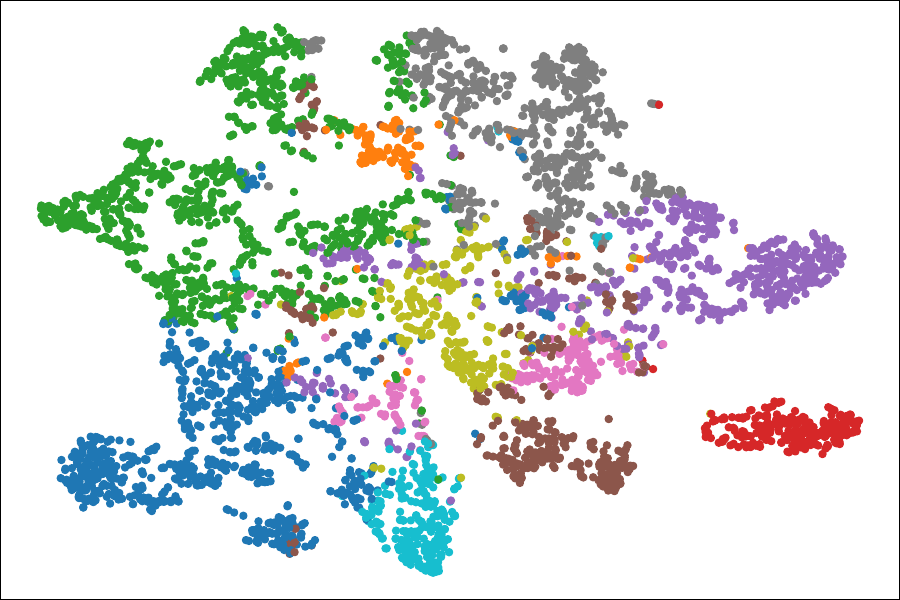}
    \label{fig:tsne_nd}
}
\subfigure[InI]{
    \includegraphics[width=0.45\linewidth]{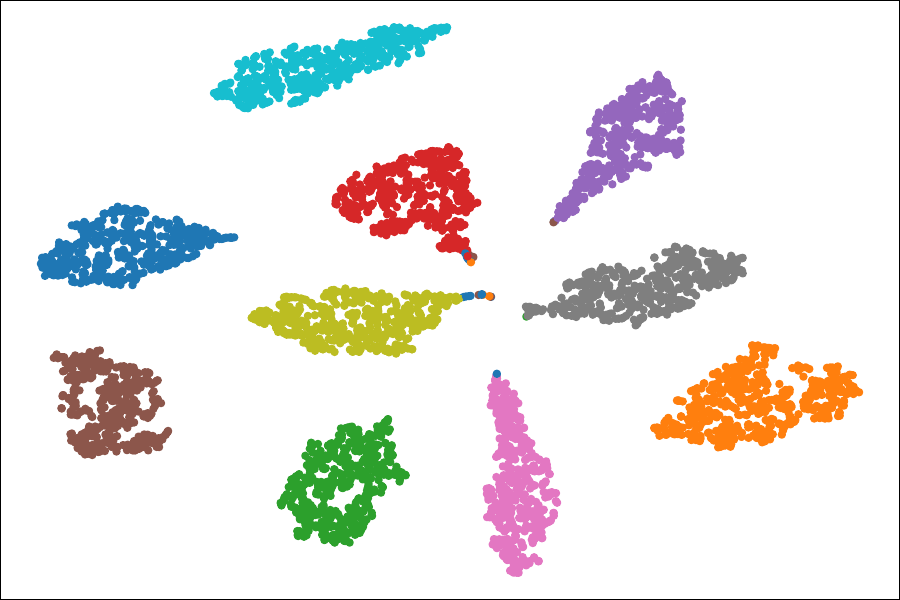}
    \includegraphics[width=0.45\linewidth]{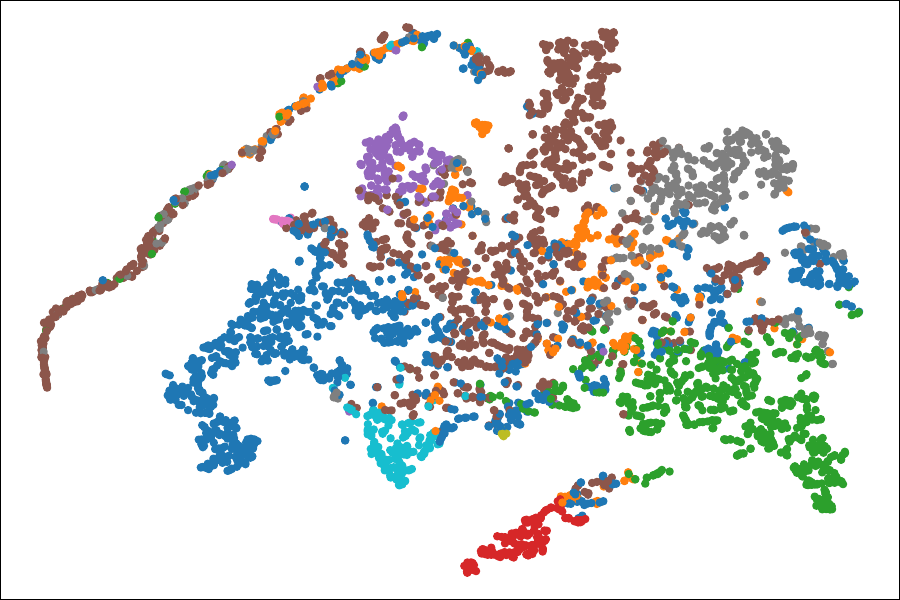}
    \label{fig:tsne_ini}
}
\end{center}
\caption{The feature visualization of undefended and defended victim models (Vanilla and InI). The first column exhibits samples from the ID dataset (MNIST), and the second column exhibits samples from the OOD dataset (EMNISTLetters). The colors of the points indicate the categories classified by the model.}
\label{fig:tsne}
\end{figure}

Figure~\ref{fig:tsne_nd} shows the feature distribution extracted by the undefended model (vanilla). Apparently, we can observe that samples from the target distribution (ID) can be clearly categorized into 10 different clusters. As for the OOD samples, they exhibit no clusters within the feature space, however, the boundaries among different classes are comparatively clear, thereby facilitating the model stealing on ODD queries. Figure~\ref{fig:tsne_ini} presents the feature distribution extracted by our InI. The clusters of ID samples remain largely unaltered, while the distribution of OOD samples shows significant differences: the boundaries between different classes become more ambiguous, which hinders model stealing attacks. We assume that the adversary induction minimizes the disparity between samples categorized in different classes, and the gradient isolation obfuscates the victim's decision boundary through the adversary's update gradient.

\section{Additional Experimental Results}
We evaluate our method on VGG-16 networks. Table~\ref{tab:benign_vgg} shows the benign accuarcy of each defense. The defense performance of KnockoffNets and JBDA are exhibited in Table~\ref{tab:knockoff_vgg} and \ref{tab:jbda_vgg}. The results are similar with those on ResNet-18, while InI achieves better defense performance against JBDA on FashionMNIST and CIFAR10.

\begin{table}[!htbp]
    \caption{Benign accuracy of each defense on different image classification datasets.}
    \label{tab:benign_vgg}
    \begin{tabular}{@{}ccccc@{}}
    \toprule
    \multirow{2}{*}{Dataset} & \multicolumn{4}{c}{Benign Accuracy}       \\ \cmidrule(l){2-5} 
                             & MNIST & FashionMNIST & CIFAR-10 & CIFAR-100 \\ \midrule
    Vanilla                  & 99.63 & 94.02        & 93.24   & 72.75    \\
    MAD                      & 99.43 & 93.84        & 92.58   & 72.04    \\
    AM                       & 99.42 & 93.63        & 92.32   & 71.48    \\
    EDM                      & 99.62 & 93.76        & 92.85   & 71.12    \\
    InI (\textbf{Ours})      & 99.45 & 93.97        & 92.39   & 72.03    \\ \bottomrule
    \end{tabular}
\end{table}

\begin{table*}[htbp]
    \caption{Experimental results for KnockoffNets attack. We report the clone accuracy and the relative performance on the target test set. Lower clone accuracy/relative performance indicates better defense performance. ``InI + MAD'' and ``InI + AM'' indicate the integration of our InI with other defenses.}
    \label{tab:knockoff_vgg}
    \begin{tabular}{@{}ccccccccc@{}}
    \toprule
    \multirow{2}{*}{Defense} & \multicolumn{2}{c}{MNIST}                     & \multicolumn{2}{c}{FashionMNIST}              & \multicolumn{2}{c}{CIFAR-10}                   & \multicolumn{2}{c}{CIFAR-100}                 \\ \cmidrule(l){2-9} 
                             & soft-label            & hard-label            & soft-label            & hard-label            & soft-label            & hard-label            & soft-label           & hard-label            \\ \midrule
    Vanilla                  & 99.52(1.00×) & 99.12(0.99×) & 64.35(0.68×) & 49.95(0.53×) & 82.35(0.88×) & 78.21(0.84×) & 35.89(0.49×) & 24.87(0.34×) \\
    MAD                      & 97.28(0.98×) & 98.81(0.99×) & 46.69(0.50×) & 43.72(0.47×) & 73.10(0.79×) & 70.17(0.76×) & 24.97(0.35×) & \textbf{20.28(0.28×)} \\
    AM                       & 98.49(0.99×) & 97.45(0.98×) & 28.18(0.30×) & 35.63(0.38×) & 74.41(0.81×) & 69.01(0.75×) & 32.95(0.46×) & 21.51(0.30×) \\
    EDM                      & 98.62(0.99×) & 97.99(0.98×) & 21.61(0.23×) & 26.08(0.28×) & 69.98(0.75×) & 68.81(0.74×) & 29.34(0.41×) & 22.49(0.32×) \\
    InI (\textbf{Ours})      & \textbf{79.90(0.80×)} & \textbf{97.27(0.98×)} & \textbf{20.42(0.22×)} & \textbf{25.57(0.27×)} & \textbf{15.17(0.16×)} & \textbf{66.36(0.72×)} & \textbf{6.63(0.09×)} & 22.39(0.31×) \\ \midrule \midrule
    InI + MAD                & 75.21(0.76×) & 97.05(0.98×) & 20.91(0.22×) & 25.69(0.27×) & \textbf{13.59(0.15×)} & \textbf{64.77(0.70×)} & 6.22(0.09×) & \textbf{13.71(0.19×)} \\
    InI + AM                 & \textbf{31.22(0.31×)} & \textbf{97.03(0.98×)} & \textbf{19.67(0.21×)} & \textbf{17.91(0.19×)} & 14.93(0.16×) & 66.76(0.72×) & \textbf{6.13(0.09×)} & 21.76(0.30×) \\ \bottomrule
    \end{tabular}
\end{table*}

\begin{table*}[htbp]
    \caption{Experimental results for JBDA attack. We report the clone accuracy and the relative performance on the target test set. Lower clone accuracy/relative performance indicates better defense performance. ``InI + MAD'' and ``InI + AM'' indicate the integration of our InI with other defenses.}
    \label{tab:jbda_vgg}
    \begin{tabular}{@{}ccccccccc@{}}
    \toprule
    \multirow{2}{*}{Defense} & \multicolumn{2}{c}{MNIST}                     & \multicolumn{2}{c}{FashionMNIST}              & \multicolumn{2}{c}{CIFAR-10}                   & \multicolumn{2}{c}{CIFAR-100}                \\ \cmidrule(l){2-9} 
                             & soft-label            & hard-label            & soft-label            & hard-label            & soft-label            & hard-label            & soft-label           & hard-label           \\ \midrule
    Vanilla                  & 76.62(0.77×) & 43.32(0.43×) & 50.60(0.54×) & 57.77(0.61×) & 13.08(0.14×) & 12.67(0.14×) & 2.60(0.04×) & 3.00(0.04×) \\
    MAD                      & 16.63(0.17×) & 29.49(0.30×) & 48.11(0.51×) & 56.48(0.60×) & 12.16(0.13×) & 12.08(0.13×) & \textbf{1.08(0.01×)} & 2.97(0.04×) \\
    AM                       & 70.03(0.70×) & 25.20(0.25×) & 66.76(0.71×) & 53.88(0.58×) & 16.09(0.17×) & 12.21(0.13×) & 1.79(0.03×) & 2.93(0.04×) \\
    EDM                      & 61.28(0.62×) & 69.79(0.70×) & 51.74(0.55×) & 49.28(0.53×) & 10.88(0.12×) & 15.59(0.17×) & 2.42(0.03×) & 2.60(0.04×) \\
    InI (\textbf{Ours})      & \textbf{10.64(0.11×)} & \textbf{24.33(0.24×)} & 1\textbf{7.21(0.18×)} & \textbf{48.96(0.52×)} & \textbf{10.38(0.11×)} & \textbf{11.76(0.13×)} & 1.78(0.02×) & \textbf{2.33(0.03×)} \\ \midrule \midrule
    InI + MAD                & \textbf{9.85(0.10×)} & \textbf{3.02(0.03×)} & 31.86(0.34×) & 49.93(0.53×) & 11.50(0.12×) & \textbf{11.57(0.13×)} & 1.74(0.02×) & \textbf{2.33(0.03×)} \\
    InI + AM                 & 11.41(0.11×) & 22.57(0.23×) & \textbf{19.60(0.21×) }& \textbf{44.35(0.47×)} & \textbf{10.31(0.11×)} & \textbf{11.57(0.13×)} & 2.02(0.03×) & 2.58(0.04×) \\ \bottomrule
    \end{tabular}
\end{table*}

\end{document}